\documentclass[pdflatex,sn-mathphys-num]{sn-jnl}

\usepackage{graphicx}%
\usepackage{multirow}%
\usepackage{amsmath,amssymb,amsfonts}%
\usepackage{amsthm}%
\usepackage{mathrsfs}%
\usepackage[title]{appendix}%
\usepackage{xcolor}%
\usepackage{textcomp}%
\usepackage{manyfoot}%
\usepackage{booktabs}%
\usepackage{algorithm}%
\usepackage{algorithmicx}%
\usepackage{algpseudocode}%
\usepackage{listings}%
\usepackage[T1]{fontenc}

\begin{document}

\title[Ideology and polarization set the agenda on social media]{Ideology and polarization set the agenda on social media}

\author*[1]{Edoardo Loru}\email{edoardo.loru@uniroma1.it}
\author[2]{Alessandro Galeazzi}
\author[3]{Anita Bonetti}
\author[4]{Emanuele Sangiorgio}
\author[5]{Niccolò Di Marco}
\author[6]{Matteo Cinelli}
\author[7]{Max Falkenberg}
\author[8, 9]{Andrea Baronchelli}
\author*[6]{Walter Quattrociocchi}\email{walter.quattrociocchi@uniroma1.it}

\affil[1]{Sapienza University of Rome, Department of Computer, Control and Management Engineering, Rome, Italy}

\affil[2]{University of Padova, Department of Mathematics, Padova, Italy}

\affil[3]{Sapienza University of Rome, Department of Communication and Social Research, Rome, Italy}

\affil[4]{Sapienza University of Rome, Department of Social Sciences and Economics, Rome, Italy}
\affil[5]{Tuscia University, Department of Legal, Social, and Educational Sciences, Viterbo, Italy}
\affil[6]{Sapienza University of Rome, Department of Computer Science, Rome, Italy}
\affil[7]{Central European University, Department of Network \& Data Science, Vienna, Austria}
\affil[8]{City St George’s, University of London, Department of Mathematics, London, UK}
\affil[9]{The Alan Turing Institute, British Library, London, UK}

\abstract{\textbf{\textcolor{red}{Please refer to published version:}} \href{https://doi.org/10.1038/s41598-025-19776-z}{10.1038/s41598-025-19776-z}\\The abundance of information on social media has reshaped public discussions, shifting attention to the mechanisms that drive online discourse. This study analyzes large-scale Twitter (now X) data from three global debates—Climate Change, COVID-19, and the Russo-Ukrainian War—to investigate the structural dynamics of engagement. Our findings reveal that discussions are not primarily shaped by specific categories of actors, such as media or activists, but by shared ideological alignment. Users consistently form polarized communities, where their ideological stance in one debate predicts their positions in others. This polarization transcends individual topics, reflecting a broader pattern of ideological divides. Furthermore, the influence of individual actors within these communities appears secondary to the reinforcing effects of selective exposure and shared narratives. Overall, our results underscore that ideological alignment, rather than actor prominence, plays a central role in structuring online discourse and shaping the spread of information in polarized environments.}

\keywords{Social Media, Ideological Alignment, Polarization, Online Discourse, Online Engagement, Public Debates}

\maketitle

\section*{Introduction}
\label{sec:introduction}

The rapid expansion of social media has fundamentally changed the way people communicate, share information, and consume content. These platforms are now central to society, serving as tools for dissemination, entertainment, and consumption of information \cite{bakshy2012role, Tucker2018, aichner2021twenty, DiMarco2024}. However, their engagement-driven business models and the dynamics of online interactions have raised concerns about their broader social impacts \cite{van2021social, guess2023social}, including their role in deepening polarization \cite{flaxman2016filter, cinelli2021echo, nyhan2023like}, propagating misinformation \cite{del2016spreading, shmargad2020sorting, gonzalez2023asymmetric}, and amplifying hate speech \cite{siegel2020online, castano2021internet, lupu2023offline, avalle2024persistent, loru2024coordinated}.
Moreover, recent research has highlighted the central role that the structure of social networks has in disseminating information \cite{Ji2023}.

Building on these concerns, today’s digital environment is characterized by an overabundance of information, offering users an unprecedented range of choices for media consumption \cite{van2017political}. This shift results from the proliferation of various sources of content, fueling intense competition for user attention \cite{anderson2012competition, webster2014marketplace, lorenz2019accelerating}, amplified by the engagement-driven strategies of social media platforms.
Consequently, the information ecosystem has undergone a significant transformation, leading to a fragmented audience willing to participate \cite{bruns20083, livingstone2013participation} but dispersed across numerous platforms and personalized feeds \cite{webster2012dynamics}. Traditional media gatekeepers, once central to public discourse, no longer unilaterally set the agenda \cite{perloff2022fifty}, diverging from previously dominant patterns of centralized media consumption \cite{prior2007post, gallup2018indicators, nic2018reuters, sangiorgio2024followers}. 

This radical transformation has led to a vast and unstructured flow of content \cite{chadwick2017hybrid} that may induce information overload \cite{bawden2020information}. Users often feel overwhelmed by the sheer volume of content when seeking information \cite{hargittai2012taming}, complicating their ability to navigate this landscape. In this information-saturated ecosystem, online users increasingly exhibit selective consumption behaviors, engaging primarily with content aligned with their existing beliefs \cite{schmidt2017anatomy, cinelli2020selective, gonzalez2023asymmetric} while ignoring opposing viewpoints \cite{bessi2015science, zollo2017debunking}. This behavior reinforces echo chambers and exacerbates polarization \cite{bennett2008new, bail2018exposure, cinelli2021echo}, in an environment where the value of news content is no longer determined solely by its intrinsic quality \cite{deuze2008changing}. News consumption has become a shared social experience, with individuals exchanging links and recommendations within their networks and treating news as a form of cultural currency \cite{singer2014user}. As a result, traditional hierarchical models of media influence struggle to explain the dynamics of public discourse in this decentralized environment \cite{mccombs2014new, barbera2019leads, galeazzi2024agenda}, where narratives and participants compete for prominence.

Previous studies have explored how social media have changed the traditional gatekeeping function of legacy media. Contemporary debates on social platforms are frequently shaped not by institutional actors, but by networked publics \cite{boyd2010social} participating in the online public sphere. In classic media research, gatekeeping was understood as the process through which mass media filtered, shaped, and prioritized information flows to audiences \cite{shoemaker1991gatekeeping, Shoemaker2001Individual}. However, the rise of participatory digital technologies has profoundly challenged this top-down model. Scholars have argued that social media has empowered users to actively contribute to the creation and dissemination of news content, thereby eroding the media’s monopoly on gatekeeping \cite{bowman2003we, gillmor2006we}.
As Bro and Wallberg\cite{Bro2014} note, individuals increasingly access news through peer networks, such as family and friends, rather than through editorially curated channels. This shift has altered not only how news spreads, but also who holds the power to define it. While news organizations continue to seek authority over informational flows \cite{Pantic2024}, alternative actors have emerged as influential newsmakers in the digital sphere. This transformation has led some scholars to reconceptualize gatekeeping as a distributed, collaborative process within communicative networks. Hoskins and O’Loughlin \cite{Hoskins2011} initially questioned whether gatekeeping could be meaningfully applied to such networked contexts, but their research—along with subsequent studies—supports the notion of “network gatekeeping,” a dynamic process emerging from diffused and collaborative gatekeeping processes \cite{Hoskins2011,Meraz2013}. Our findings contribute to this literature by illustrating how agenda-setting processes might be increasingly negotiated among a wide array of actors, in an environment where news also emerges from bottom-up dynamics.

While classical agenda-setting theory, originating with McCombs and Shaw \cite{mccombs1972the}, was built around a top-down logic typical of mass communication, subsequent scholarship has explored its evolution in light of media digitalization, debating the relevance of the theory in digital settings. 
The same authors, on the 50th anniversary of the theory, traced its evolution by identifying seven distinct facets, emphasizing need for orientation, network agenda setting, and agendamelding as particularly active areas of contemporary research \cite{shaw1999agenda, shaw2019agendamelding} within an increasingly complex and interactive media ecosystem \cite{mccombs2018new}.  Some scholars have argued that agenda setting effects remain operative, albeit adapted to the new media landscape \cite{roberts2002agenda}, while others have pointed to a weakening of traditional media's influence in favor of digital platforms \cite{wallsten2007agenda, meraz2011using}, or have conceptualized social media themselves as a further development of agenda setting mechanisms \cite{johnson2013agenda}. Russell Neuman et al. \cite{russell2014dynamics} identified several patterns of issue framing in social and traditional media that highlighted an interdependence of the two rather than a one-way agenda-setting dynamic. Feezell \cite{feezell2018agenda}, instead, investigated how social media contributes to the dissemination of mainstream news and their perceived importance. New studies on agenda setting processes have also kept in consideration various platforms, such as Twitter \cite{ceron2014twitter, su2019agenda, rogstad2016twitter}, Instagram \cite{towner2024instagram} and Reddit \cite{kim2025agenda}. Analyzing 71.77 million tweets, Yi and Wang \cite{yi2022affecting} found that professional media influence only a small number of individuals, while they argue that opinion leaders have become strong competitors of traditional professional media in agenda setting processes.

Understanding these processes requires a change in perspective. Using multiple large-scale Twitter (now X) datasets, our analysis outlines the drivers of public discourse by examining three global debates—climate change (COP26), COVID-19, and the Russo-Ukrainian War. We compare various actor types, such as media outlets, activists, and individual influencers, to determine whether specific categories dominate discussions and what drives online debates. Furthermore, our diverse selection of debates, spanning environmental, health, and geopolitical domains, enables us to assess whether the dynamics we investigate are tied to specific topics or reflect systemic patterns across issues.

In particular, we aim to address the following questions: What are the drivers of user engagement in fragmented social media environments? Do actor categories or ideological alignment better explain how attention is allocated?
Our findings reveal a fragmented environment where no single category of actor consistently dominates discussions. Furthermore, users cluster into two polarized ideological communities, one representing the majority of users endorsing a mainstream narrative and a minority usually embracing alternative views. Notably, a user's stance in one debate is a reliable marker of their opinion on other topics. For instance, users supporting climate action in the COP26 debate are likely to advocate for vaccination during COVID-19 and aid for Ukraine, while opponents of COVID-19 vaccination are typically climate skeptics and critics of military aid. This consistency in ideological stance across heterogeneous debates suggests that polarization on social media emerges from overarching ideological divides, rather than being confined to specific issues.

Our results underscore that online discussions are shaped by shared ideological alignments, rather than by a top-down model dominated by traditional gatekeepers. This highlights how decentralized, bottom-up dynamics dominate public discourse, where ideological structures are pivotal in predicting the evolution of debates and the spread of specific narratives. By identifying consistent patterns in user engagement and ideological alignment, this study provides insights into the mechanisms shaping public discourse in an information-saturated landscape.

\section*{Results and Discussion}
\label{sec:results}

We investigate the relationship between engagement and ideological configurations across three major debates---COP26, COVID-19, and the Russo-Ukrainian War---employing data collected with specific keyword searches (see Methods). We begin by analyzing the influence of various types of actors in driving online discussions. Next, we quantify the degree of polarization surrounding different topics and examine the relationships between users' opinions across these topics. Finally, we explore how users in different ideological communities interact with various types of actors across topics.

\subsection*{Decentralized Engagement: Beyond Influencer Categories}
\label{sec:results_rq1}

To investigate the dynamics of engagement in online discussions, we identify a set of accounts that we refer to as `influencers'. We first select the top 1\% of accounts receiving the highest number of retweets within each debate. We further refine the selection from this subset to include only the top 50\% producers of original tweets. This approach ensures that our analysis focuses on accounts that receive significant attention and contribute to the discussion by actively creating new content.

Each influencer is then manually assigned to one of six categories: Activist, International Organization/NGO, Media, Politics, Private Individual, and Other. Accounts whose tweets are unrelated to the topics of discussion are excluded from the analysis. In addition to categorizing these accounts according to their role, we annotate each account with their stance on key issues within the debates. The final sets of influencers are comprised of 561 accounts for COP26, 1577 for COVID-19, and 1166 for Ukraine. The labeling process is described more in detail in Methods and in Supplementary Information, where we also report the proportion of influencers in each category (Supplementary Fig. S1) and the distributions of original tweets posted by these accounts (Supplementary Fig. S2).

One fundamental question is the following: according to agenda-setting theory, we might expect media or political accounts to receive higher levels of engagement than other types of information providers. To understand if this is the case, namely whether the engagement that influencers receive depends on the category they belong to, we analyze the distribution of retweets received by each category per debate. This analysis enables us to assess whether certain categories of influencers are more effective in driving users' engagement.

\begin{figure}[!t]
\centering
\includegraphics[width=1\textwidth]{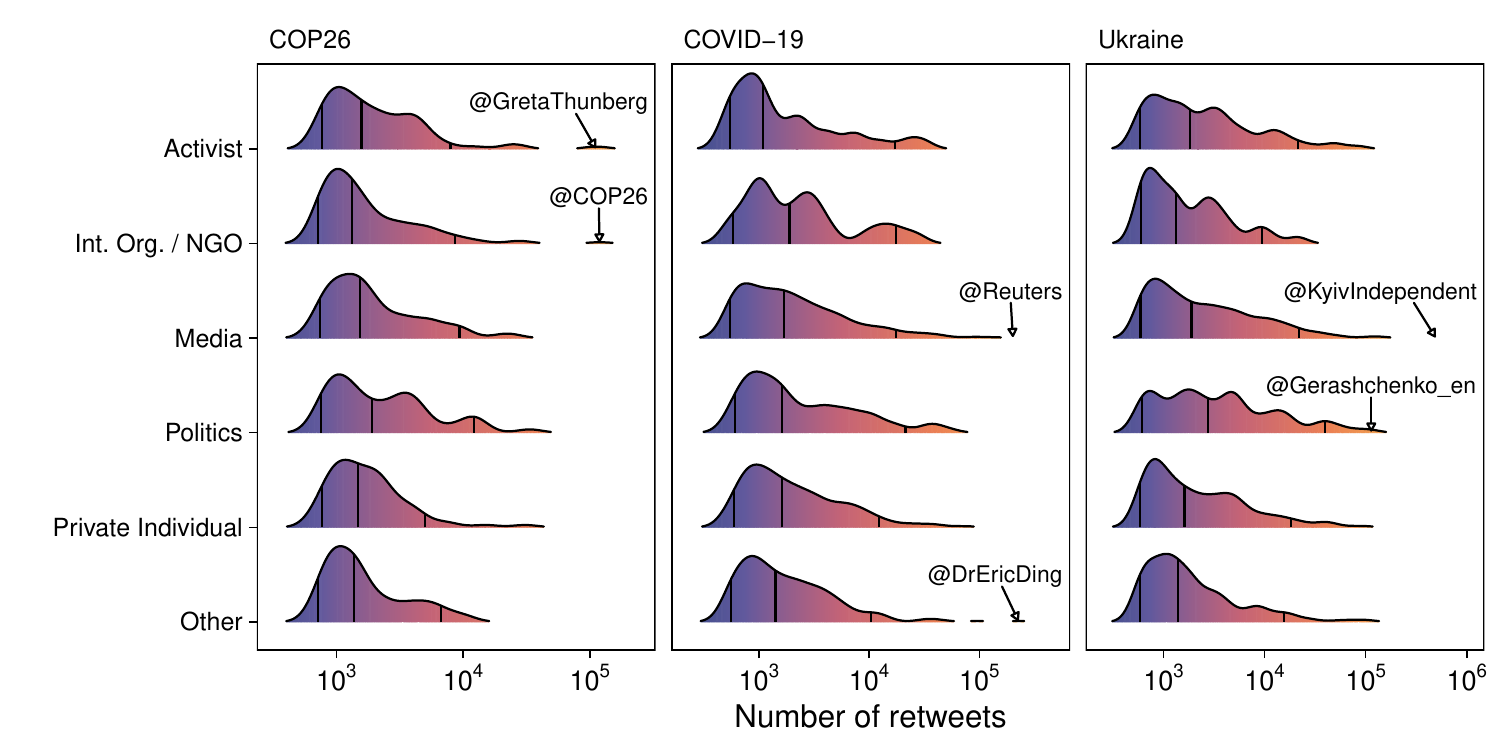}
\caption{\textbf{Distribution of the number of retweets received by each influencer category and for each debate.} The vertical lines reported over each distribution indicate, from left to right, the 0.05, 0.5, and 0.95 quantiles. We also report the usernames of the two most retweeted accounts per debate. No category appears to dominate over the others, but rather singular prominent accounts in each debate emerge over the rest. Notable examples are @GretaThunberg in the climate debate and @KyivIndependent in the debate about the Russo-Ukrainian War.}
\label{fig:influencers_pdf_retweets}
\end{figure}

The distributions displayed in Fig. \ref{fig:influencers_pdf_retweets} do not indicate that specific categories of actors consistently receive more retweets. Instead, all categories receive comparable attention, suggesting that the category itself may not be a determining factor for the volume of retweets. For instance, although \texttt{@GretaThunberg} stands out as a prominent figure in the COP26 discussion, other activists are retweeted with comparable frequency to influencers from other categories. Similarly, in the Ukraine debate, accounts in the Media category attract engagement levels similar to influencers from the other categories, with few exceptions such as \texttt{@KyivIndependent}---a team of independent journalists.

To statistically test these insights, we perform pairwise comparisons between the categories within each debate using Kolmogorov-Smirnov tests, which assess whether the retweet distributions for any two categories are drawn from the same population. The results, reported in Supplementary Table S1, show no statistically significant differences between distributions, suggesting that the category of an influencer does not affect the number of retweets they receive. Additional results are provided in Supplementary Information, including the prevalence of each influencer category across retweet count buckets (Supplementary Fig. S3).

To further explore the role of influencer categories, we test whether the retweet distributions align with a null model in which these labels are assigned randomly. Two randomization strategies are applied:

\begin{enumerate}
    \item \textbf{Reshuffling}: Category labels are randomly reshuffled across influencers, to ensure each category keeps its original size.
    \item \textbf{Uniform Sampling}: Categories are randomly sampled from a uniform distribution, with each having an equal probability of selection ($p = 1/6$).
\end{enumerate}

We compare the observed and randomized distributions using two-sided two-sample Wilcoxon rank-sum tests for both strategies. The results show no significant differences, indicating that the observed number of retweets is consistent with random assignment of categories. This reinforces the finding that an actor's category does not appear to affect their ability to attract retweets. Detailed results are provided in Supplementary Table S2, as well as visualizations of the randomized distributions (Supplementary Fig. S4).

Finally, we analyze users' preferences in retweeting influencer categories. To ensure that all users in a dataset might have retweeted each category at least once, we restrict our analysis to users with at least six retweets and compute how frequently they retweet each category. We then employ the Gini Index (GI)~\cite{gini1921measurement} to measure how evenly users distribute their retweets across influencer categories. The GI is a value in $[0,1]$ that quantifies inequality within a distribution: a value of $0$ corresponds to minimum inequality (i.e., all categories are retweeted by the user with the same frequency), while a value of $1$ corresponds to maximum inequality (i.e., the user only retweets a single category). See Methods for its formal definition.

The resulting distributions of users' GI, displayed in Supplementary Fig. S5, suggest that users show a preference for a limited subset of categories. For COP26, the GI distribution has a median of $0.73$ (interquartile range: $Q1 = 0.60,\ Q3 = 0.87$); for COVID-19, the median is $0.77$ ($Q1 = 0.65,\ Q3 = 0.89$); and for Ukraine, the median is $0.69$ ($Q1 = 0.58,\ Q3 = 0.80$). However, when considering the total number of retweets each category receives, we observe lower levels of inequality across all debates: $\text{GI} = 0.28$ for COP26, $\text{GI} = 0.59$ for COVID-19, and $\text{GI} = 0.53$ for Ukraine.

The difference between the GI estimates at the user and debate level indicates that, while most users tend to favor certain categories over others, the specific subset of preferred categories varies across individuals. Users retweet different influencer categories, but these individual preferences balance out, resulting in smaller overall inequality across categories. This pattern is visible in all three debates, particularly in the case of COP26.

\subsection*{Ideological Persistence as a Driver of Engagement}
\label{sec:results_rq2}

Our analyses show that considering only influencers' categories is insufficient to explain the drivers of user engagement on social media. As other factors such as ideological alignment~\cite{falkenberg2022growing} have been shown to influence how people interact with social media content, we now investigate the role of ideology in shaping engagement across the key debates analyzed in this study.

In line with previous research \cite{barbera2015tweeting, flamino2023political, falkenberg2022growing}, we estimate the ideological leaning of an individual $i$ using the ``latent ideology'' method. This technique infers the ideological distribution within each debate by assigning each individual a scalar value $x_i$ representing their ideological position. These values are derived from the users' interactions with influencers, allowing us to capture potential patterns of ideological polarization. 

We begin by constructing an interaction matrix where each element corresponds to the number of times a user retweets a specific influencer. Then, the method assigns numerical scores to users so that individuals retweeting similar subsets of influencers receive similar scores. By rescaling these values to the range $[-1, +1]$, we obtain an ideological spectrum, with users at opposing ends interacting with distinct and non-overlapping sets of influencers. A detailed explanation of the technique is provided in Methods.

Following the methodology adopted in previous works \cite{falkenberg2022growing}, we focus on the top 300 most retweeted influencers from each of the three debates. We calculate the latent ideology of all users within each debate and validate the resulting scores to ensure that the distributions represent true ideological spectra (see Supplementary Section 2.1 and Supplementary Fig. S6). Finally, we follow standard procedure and estimate the ideological position of each influencer as the median ideology score of their retweeters \cite{falkenberg2022growing}.

\begin{figure}[t!]
\centering
\includegraphics[width=1\textwidth]{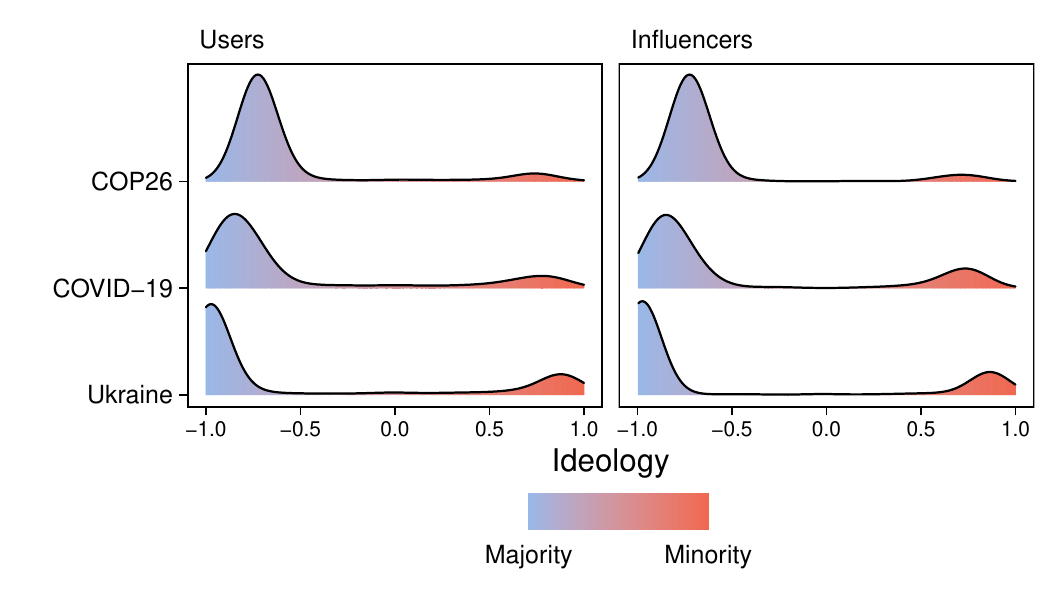}
\caption{\textbf{Ideological spectra of users and influencers in the COP26, COVID-19, and Ukraine debates.} The majority group is mapped to $-1$, whereas the minority group to $+1$. All distributions exhibit bimodality, indicating the presence in each debate of two main communities of users who retweet distinct sets of influencers.}
\label{fig:ideology_pdf}
\end{figure}

Figure \ref{fig:ideology_pdf} illustrates the latent ideology distributions for users and influencers across the three debates. To ensure consistency, we map the majority group to $-1$ and the minority group to $+1$ for all cases. Although the extent of polarization varies between debates, all distributions exhibit a clear bimodal structure, with most users clustered near the spectrum's extremes. This observation indicates the existence of two dominant, ideologically opposing communities within each debate.

To formally evaluate the previous observation, we compute Hartigan's dip statistic to test for unimodality (see Methods). The results are consistent across debates, with $D = 0.021$ $(P < 0.001)$ for COP26, $D = 0.035$ $(P < 0.001)$ for COVID-19, and $D = 0.068$ $(P < 0.001)$ for Ukraine. These values indicate a statistically significant deviation from unimodality, confirming the presence of two well-separated communities in each debate. These ideologically opposing clusters reflect the presence of two polarized communities, where users predominantly interact with influencers aligned with their ideological stance.

Having manually annotated all influencers, we can identify the ideological stances most representative of these communities. Influencers in the Majority predominantly advocate for pro-climate action, pro-vaccination, and pro-aid stances in the COP26, COVID-19, and Ukraine discussions, respectively. Conversely, influencers in the Minority are typically associated with alternative or fringe viewpoints, such as skepticism toward climate change and COVID-19 vaccination. For instance, accounts with large followings such as \texttt{@JunkScience} for COP26, \texttt{@HighWireTalk} for COVID-19, and \texttt{@georgegalloway} for Ukraine, were all assigned to the ideological Minority.

Consistent with previous research on online polarization \cite{garimella2017quantifying, cota2019quantifying, cinelli2021echo, falkenberg2022growing}, our findings reveal the presence of communities supporting conflicting ideologies. This dynamic fosters user engagement with one of these opposing narratives, resulting in a recurring pattern of polarization across debates.

\begin{figure}[t!]
\centering
\includegraphics[width=1\textwidth]{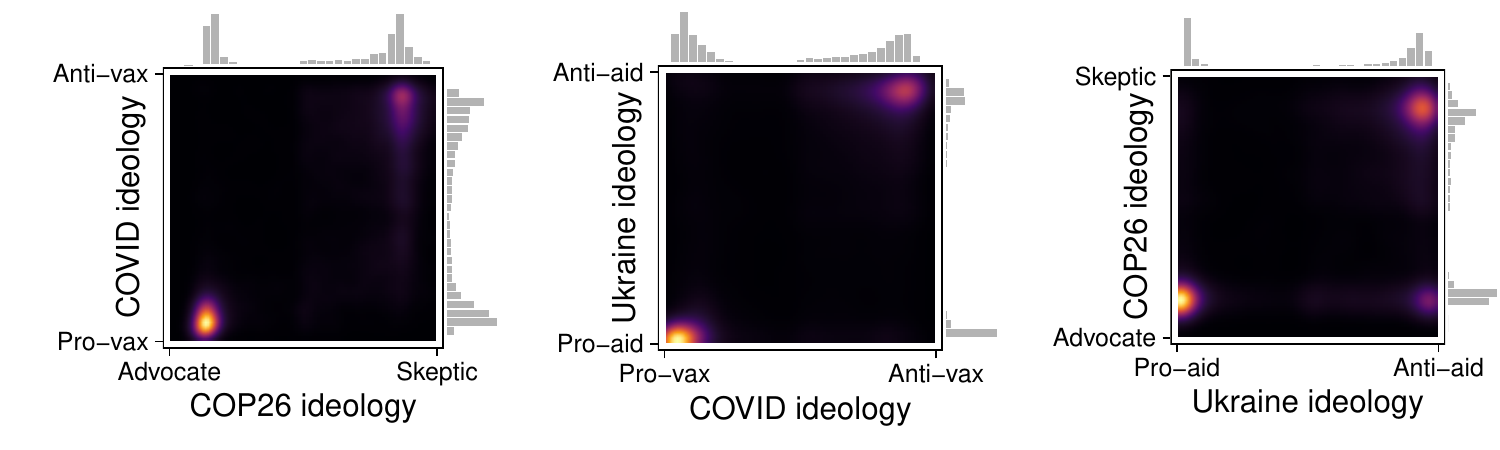}
\caption{\textbf{Joint ideology spectra of users for each pair of debates, with marginal distributions on the top and right sides.} The `Majority' (`Minority') ideology is mapped to the left (right) of the $x$-axis and the bottom (top) of the $y$-axis. Colors indicate the density of users in a region, with lighter colors corresponding to a higher concentration. Most users occupy the regions on the bottom-left and the top-right of the ideology space, indicating a substantial overlap between the Majority and Minority communities across debates.}
\label{fig:ideology_joint}
\end{figure}

To evaluate the persistence of ideological segregation between 
different topics, we quantify the ideological overlap of users across debates. Figure \ref{fig:ideology_joint} presents the density plot of the joint distribution $P(x,y)$ of users' ideologies, where $P(x)$ and $P(y)$ correspond to the ideology distributions for the debates represented on the $x$ and $y$ axes, respectively. This analysis involves $N=43,392$ users with an ideology score both in the COP26 and COVID-19 debates, $N=91,647$ in COVID-19 and Ukraine, and $N=28,463$ in Ukraine and COP26.

In the figure, points along the bisector of the first and third quadrants represent users who maintain a consistent ideological stance (Majority or Minority) across both debates. Conversely, points along the bisector of the second and fourth quadrants represent users with misaligned stances between debates (e.g., Majority in one debate and Minority in the other). 
To account for the size imbalance between the Majority and Minority groups shown in Fig. \ref{fig:ideology_pdf}, we generate each panel of Fig. \ref{fig:ideology_joint} by first downsampling $P(x)$ and then $P(y)$, to ensure that the number of users in the Majority group matches that of the Minority group. The full distributions without downsampling are provided in Supplementary Fig. S7.

Our findings reveal a substantial overlap of ideological communities across debates. This is reflected by the high density of users in the bottom-left (Majority in both) and top-right (Minority in both) corners of the joint distributions. Interestingly, the joint distribution of Ukraine and COP26 ideologies reveals a subset of users who hold a minority, anti-aid stance while simultaneously supporting climate action, such as \texttt{@ReadeAlexandra} and \texttt{@DeborahDupre}. This divergence suggests that the strength of ideological alignment may vary depending on the topics under discussion. 

To further assess the persistence of the Majority versus Minority configuration across debates, we calculate the probability of a user maintaining the same ideological stance across different topics. Specifically, we estimate the fraction of users within the Majority or Minority group in one debate who belong to the same group in the other two debates.

\begin{figure}[t!]
\centering
\includegraphics[width=1\textwidth]{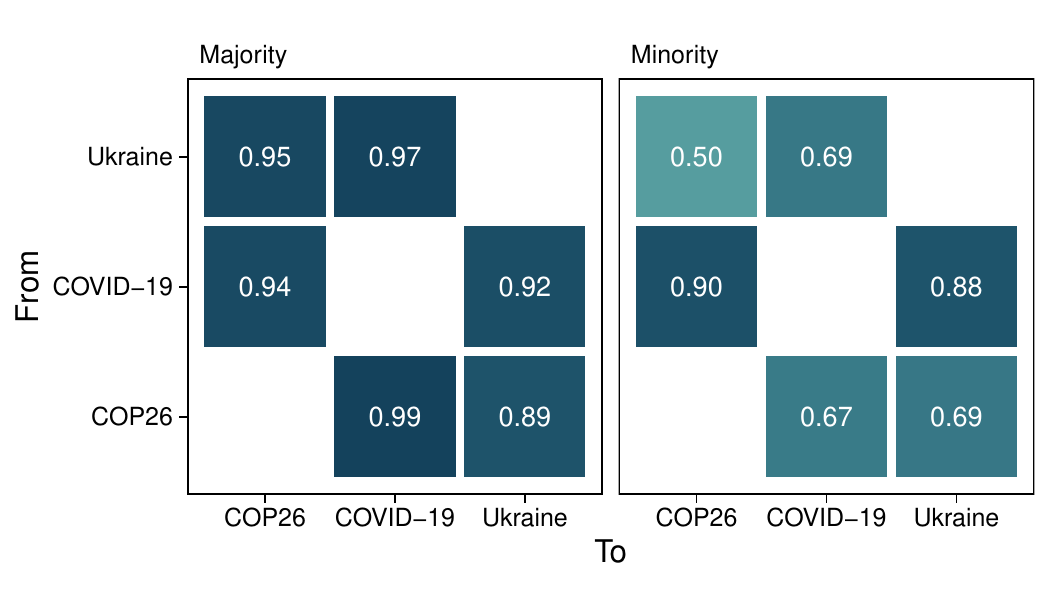}
\caption{\textbf{Conditional probability of user ideology across debates.} Each tile reports the fraction of users in a `From' debate who belong to the same community in a `To' debate. More than 90\% of users with `Majority' ideology on a debate (left panel) stay in the majority community of the other two. Similarly, but to a lesser extent, users with `Minority' ideology on a debate (right panel) tend to stay in the minority community of the other two, especially users within the ideological minority of the COVID-19 debate.}
\label{fig:cond_prob}
\end{figure}

The results, presented in Fig. \ref{fig:cond_prob}, further support the qualitative insights derived from the joint distribution in Fig. \ref{fig:ideology_joint}.
The two matrices display the fraction of users who remain in the majority (or minority) group in a debate on the $x$-axis, given that they belong to the majority (or minority) group in the debate specified on the $y$-axis. This asymmetry arises from the different fractions of majority and minority supporters in each debate. For instance, the proportion of COP26 supporters who also endorse COVID-19 vaccines may differ from the proportion of vaccine supporters who also back climate change policies. The matrices reveal that a user’s stance in one debate is highly indicative of their stance in others, with the proportion of users maintaining the same ideological leaning equaling or exceeding 90\% for most debate pairs. This result is robust to using more restrictive ideology thresholds to label users as `Majority' or `Minority' (Supplementary Fig. S8). 

Users in the Majority group of one debate are highly likely to belong to the Majority group in another. For instance, 99\% of users in the ideological Majority of the COP26 debate also belong to the Majority in the COVID-19 debate. Similar patterns are observed for users in the Minority group, albeit with slightly less consistency. Most users in the Minority of one debate tend to adopt a Minority stance in the others. Notably, the Minority group in the COVID-19 debate exhibits a high level of ideological cohesion, with approximately 90\% of these users also holding a Minority position in the COP26 and Ukraine debates. In contrast, the Ukrainian debate stands out as the only case where a Minority stance does not strongly indicate corresponding positions in both other debates.

This discrepancy may be attributed to the distinct characteristics and perceived salience of the debates under study. The COVID-19 pandemic, as a global and prolonged event, likely fostered a cohesive ideological identity among users skeptical of institutional narratives. This shared identity appears to extend across debates, such as COP26, reflecting a broader distrust of mainstream scientific and institutional messaging. Conversely, the Russo-Ukrainian War, while significant, represents a geographically and politically localized event. This localized nature may result in less ideological cohesion among Minority users across debates. Additionally, the Ukraine debate’s complexity— encompassing geopolitical perspectives, humanitarian concerns, and diverse narratives—could contribute to weaker alignment with stances on other topics.

On social media, ideological alignment has a greater impact on shaping discussions than the social, political, or professional roles of individuals or organizations. No single category of actors consistently drives the discussion. Instead, our findings highlight the persistence of a Majority versus Minority ideological configuration across debates. Users tend to consistently align with one side of the spectrum, suggesting that the structure of online discourse is less about the hierarchy of different types of actors and more about the alignment of narratives with pre-existing ideological stances.

\subsection*{Influencer Engagement Patterns in Majority and Minority Communities}

In the previous sections, we show that most influencers receive a similar number of retweets, irrespective of their category. Furthermore, users exhibit consistent behavior by retweeting influencers aligned with the Majority or Minority ideological communities across topics. Now, we aim to investigate further how the two communities differ in terms of the influencers they retweet. 

To begin, we analyze whether the number of retweets an influencer receives varies depending on the ideological community they belong to. For this purpose, we employ a bootstrapping procedure to estimate the median number of retweets and retweeters attracted by influencers associated with the Majority and Minority communities. The results, presented in Fig. \ref{fig:category_ideology_boot} and detailed in Supplementary Table S3, show that across all three debates, the estimates for the two groups of influencers are comparable. This indicates that individual influencers attract similar numbers of retweets and unique retweeters, regardless of their ideological alignment, and suggests that the ideological community to which an influencer belongs does not substantially impact the engagement they receive. 

\begin{figure}[t]
\centering
\includegraphics[width=1\textwidth]{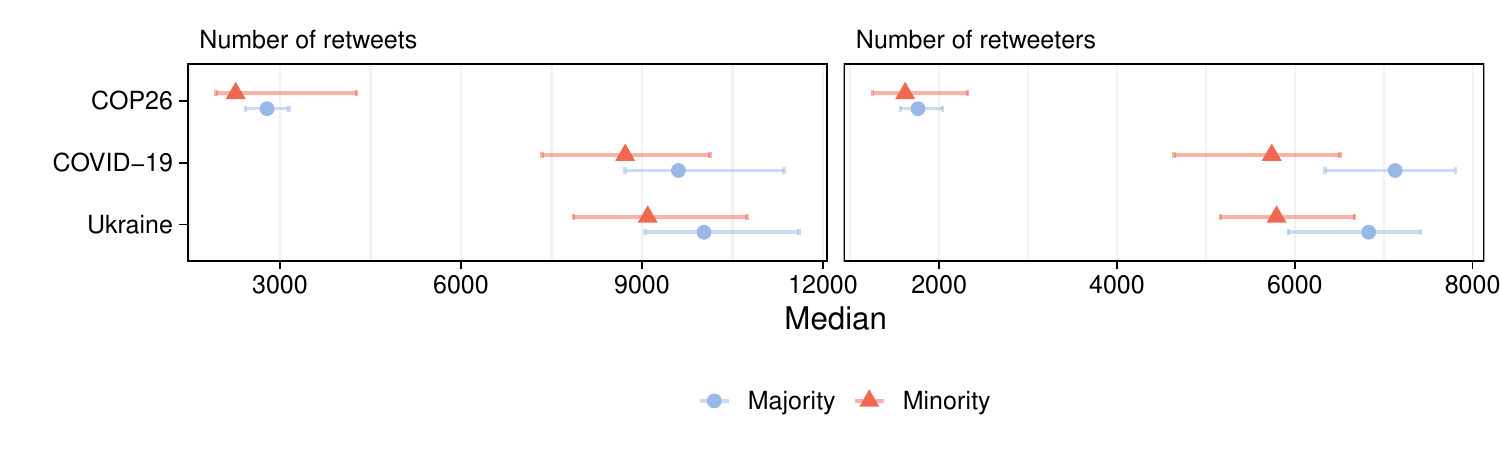}
\caption{\textbf{Median number of retweets and retweeters attracted by influencers of the majority and minority communities}. The estimates result from a bootstrapping procedure, with error bars corresponding to the 95\% bias-corrected and accelerated confidence intervals. Across all debates, the two sets of influencers receive a comparable number of retweets from a similar number of unique users.}
\label{fig:category_ideology_boot}
\end{figure}

\begin{figure}[t!]
\centering
\includegraphics[width=1\textwidth]{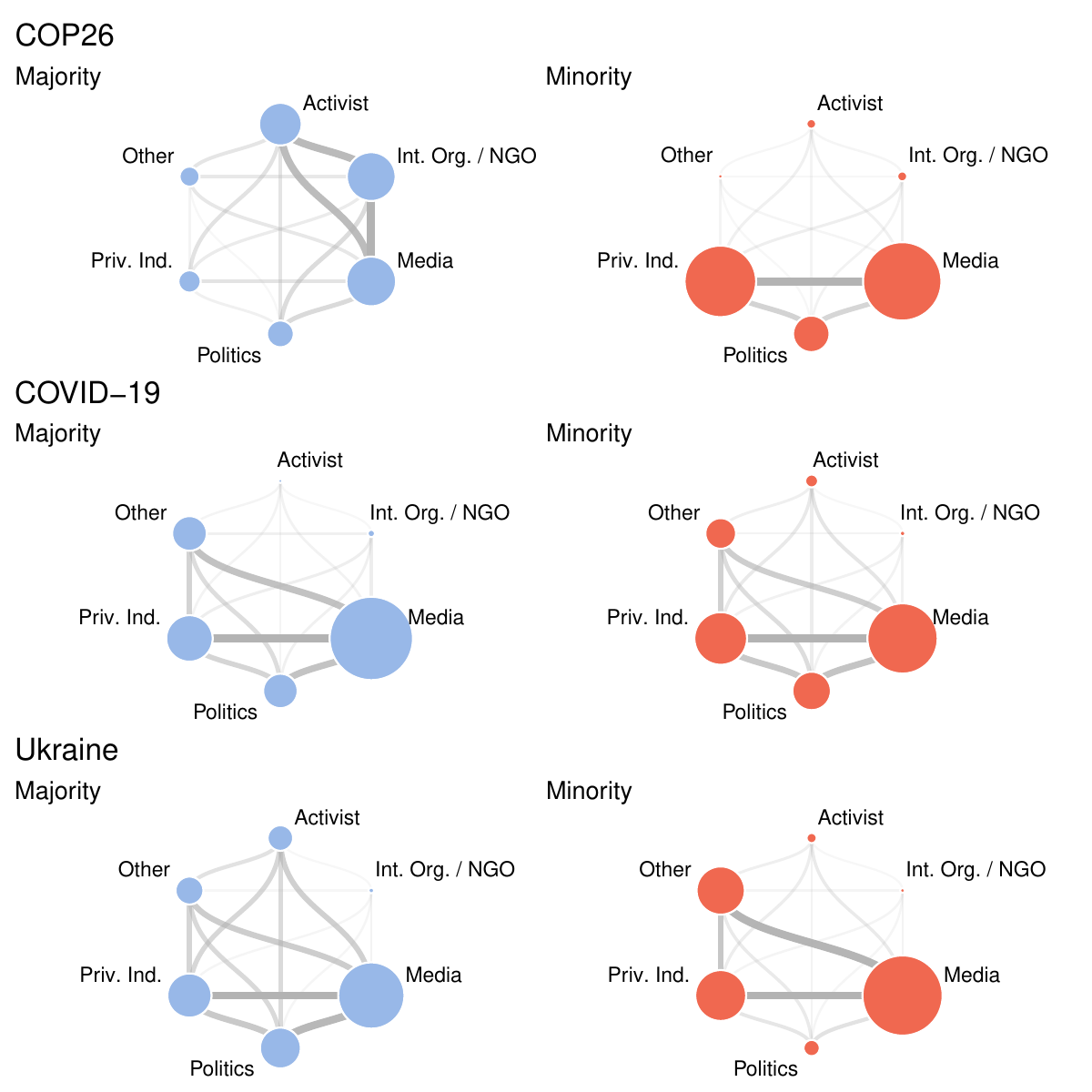}
\caption{\textbf{Networks of influencer categories for each debate.} The size of a node is proportional to the number of users who retweeted that category. The width and visibility of an edge between two categories is proportional to the number of users who retweeted both. Only retweets by users in the Majority are considered on the left column, whereas on the right column, only those by users in the Minority. The resulting networks show that the users of the two communities tend to retweet different categories of influencers. This is particularly evident in the case of the COP26 and Ukraine debates, whereas differences are not as apparent in the case of COVID-19.}
\label{fig:category_network}
\end{figure}

Next, we investigate which categories are more frequently retweeted by users with a Majority or Minority stance. For this purpose, we construct a network of influencer categories for each debate and ideological stance. In these networks, two categories are connected if a user retweets influencers from both categories at least once, with the weight of each connection representing the number of distinct users who retweeted both. Figure \ref{fig:category_network} illustrates the networks generated from this procedure.

In general, the resulting networks indicate that users in the Majority and Minority communities tend to retweet influencers from distinct categories. This distinction is particularly clear in the COP26 debate, where $35\%$ of all retweets to influencers by users in the Minority are received by Private Individuals, while the Majority primarily engages with Activists ($21\%$) or Organizations ($28\%$). Additionally, Media accounts receive a higher relative number of retweets from users in the Minority ($35\%$) compared to those in the Majority ($23\%$), reflecting a distinction in how the two ideological communities interact with information sources.

Similar patterns emerge in the Ukraine debate, albeit less pronounced. Media and Private Individuals are the more popular influencer categories for both ideological communities, accounting for $68\%$ and $71\%$ of all retweets to influencers by Majority and Minority users, respectively. In contrast, other categories are more popular in one community than the other, such as Politics ($15\%$ of total retweets by the Majority and $5\%$ by the Minority). 

Interestingly, users in the Majority and Minority of the COVID-19 debate display similar preferences, as both communities tend to retweet each category with similar frequency relative to one another. This lack of distinction indicates that no influencer category is favored more by one ideological community over the other, as we observe instead in other debates.

These findings further reinforce the conclusion that users’ engagement on social media is shaped more by ideological alignment than by influencers' professional or social roles. This dynamic reflects the polarization observed in online discourse, where individuals selectively interact with content that aligns with their preexisting views.

\section*{Conclusions}
\label{sec:discussion}

Our study provides new insights into the interplay between user engagement and influencers' roles in major global debates on social media. By analyzing large-scale datasets from the COP26, COVID-19, and Russo-Ukrainian War debates, we demonstrate that user ideological alignment holds greater significance than an influencer's category in shaping online discussions.

In traditional media environments, agenda-setting theories emphasize the hierarchical influence of institutional actors, such as mainstream media and political elites, in prioritizing public issues. Our findings suggest that these dynamics may not fully extend to social media. While certain users achieve prominence within specific topics, their relevance does not appear to be tied to their social or professional role, nor is it consistent across debates. Rather, the engagement they receive may arise from resonance within polarized communities. 

These findings hint at a change in the traditional top-down agenda setting model, pointing to a growing role for decentralized, bottom-up mechanisms, where users collectively amplify narratives aligned with their ideological preferences. The persistence of a Majority versus Minority ideological configuration in all debates highlights the presence of segregated and ideologically contrasting communities that extend beyond specific topics. This consistent affiliation may further limit exposure to diverse perspectives, thereby reinforcing the effects of echo chambers. Further, despite the difference in the size of their communities, the comparable engagement levels obtained by Majority and Minority influencers suggest that their popularity is not directly related to the ideological stance they assume.

While ideological alignment is a unifying factor, our analysis also reveals topic-specific variations. For example, the Ukraine debate exhibits weaker ideological cohesion compared to COP26 and COVID-19, particularly among Minority users. This difference likely reflects the varying salience and scope of these topics. The COVID-19 pandemic, as a global and prolonged crisis, appears to have fostered a cohesive ideological identity among users with alternative views, extending across debates such as climate change. In contrast, the Ukraine conflict, characterized by regionally and politically fragmented narratives, contributes to weaker ideological alignment across topics.

Our results also offer some noteworthy applications. 
For communication professionals, they suggest a need to rethink outreach strategies: rather than targeting audiences solely based on topic relevance or demographic features, it may be more effective to account for users’ broader ideological positioning. Tailoring messages that bridge ideological narratives or are endorsed by trusted influencers within specific communities could increase the chance of cross-cluster diffusion.
However, the susceptibility of this type of dynamic shall be taken into account also from a policymaking point of view, as it may fuel the cross-spreading of radicalized narratives between different sensitive topics. 
Additionally, our findings contribute to the understanding of filter bubbles by demonstrating that they are not isolated to specific issues but reflect systemic, cross-topic ideological alignment. This presents challenges for initiatives aimed at promoting exposure to diverse perspectives. Efforts to mitigate filter bubbles might benefit from designing interventions that operate across topics and target underlying ideological structures, rather than focusing narrowly on specific debates.

The decentralized nature of social media discourse fostered a new model of information spreading, challenging the traditional gatekeeping power of dominant actors. This new paradigm presents both opportunities and risks. On one hand, it democratizes the flow of information, amplifying diverse perspectives. On the other hand, it risks exacerbating polarization and misinformation, as users selectively engage with ideologically congruent content across topics. Moreover, as the traditional filters that once ensured a baseline of editorial standards have weakened, questions arise about the quality of content consumed in online environments. In turn, the fragmentation of the information space undermines the conditions for a shared public debate, an essential foundation of any democratic society.

Despite the robustness of our analyses, several limitations must be acknowledged. First, our investigation focuses exclusively on Twitter and may not fully reflect user dynamics on other platforms, as platform-specific biases and algorithmic curation could, in principle, influence the patterns we observe. 
Additionally, the classification of users into Majority and Minority groups relies on latent ideology modeling, which, while effective, may simplify the multifaceted nature of user beliefs.

At the same time, we only focused on English tweets, which has to be addressed as a constraint on the generalizability of the findings. However, this detail also underscores a solidity in our findings. Despite focusing on a language used predominantly by non-local audiences, we still observe strong patterns of ideological alignment and cross-topic coherence. For instance, regarding the Russo-Ukrainian debate, even within English-speaking communities—often external to the conflict zone—polarized narratives persist and align with ideological stances in other unrelated debates (e.g., climate change and COVID-19). This upshot suggests that the ideological dynamics we identify are not only local or topic-specific but reflect a broader structure of online polarization that transcends geographic and linguistic boundaries. Nevertheless, we acknowledge that expanding future analyses to include multilingual datasets could capture more nuances in local perspectives and offer a more comprehensive view of discourse dynamics within involved regions.

Future research should extend these analyses to explore cross-platform dynamics, temporal shifts in ideological alignment, and the role of emerging fringe platforms in shaping discourse.

\section*{Methods}
\label{sec:methods}

\subsection*{Data}
\label{sec:methods_data}
All data used in this work were collected via the former Twitter API for academic research, using search queries specific to each dataset. We intentionally selected three heterogeneous debates---climate change (COP26), the COVID-19 pandemic, and the Russo-Ukrainian War---due to their global relevance and high level of online engagement. Although they vary significantly in nature and content, all three debates are ideologically charged and politically salient. This selection allowed us to test whether the mechanisms we identified, including ideological polarization, are structural rather than topic-specific.

\paragraph{COP26} Tweets and user information related to the COP26 debate were collected using the `cop26' search query, from 1 June 2021 to 14 November 2021. The dataset includes approximately 8 million English tweets by 1 million distinct users.

\paragraph{COVID-19}
Data were collected from Twitter in the period from 1 January 2020 to 30 April 2021 using the official Twitter API through keyword search. We collected every publicly available tweet that contained one of the following keywords: vaccin*, dose*, pharma*, immun*, no-vax*, novax*,
pro-vax*, provax*, antivax*, anti-vax*. The final dataset results in approximately 35 million English tweets made by 8 million unique users.

\paragraph{Ukraine} Tweets related to the Russo-Ukrainian conflict were gathered by tracking specific keywords and accounts from 22 February 2022 to 17 February 2023. We leveraged a publicly available dataset \cite{Chen2023ukraine} and hydrated a random sample comprising 25\% of all available tweet IDs (\href{https://github.com/echen102/ukraine-russia}{https://github.com/echen102/ukraine-russia}). The final hydrated dataset includes approximately 85 million English tweets by 7 million distinct users.

\subsection*{Identifying the Influencers in a Debate}
\label{sec:methods_influencers}
\subsubsection*{Account Selection}
\label{sec:methods_influencers_selection}
For each dataset, we identify a set of `influencers' that shape discussions within the context of a debate. We note that, by focusing only on English tweets, the vast majority of these influential accounts will also be English-speaking. 

We start by identifying the users among a dataset's top 1\% most retweeted accounts. Users whose content is extensively reshared may have a role in driving discussions on social media, as retweeting may increase the visibility of certain content and, in turn, heighten its perceived salience in the public eye. However, focusing solely on retweets may also yield accounts that have received many retweets, but only on a small number of tweets that may have gone viral. Instead, we are interested in users who have been consistently influential in the debate, by not only receiving many retweets but also producing a sufficiently large quantity of original content. For this reason, we restrict the resulting sets of influencers in each dataset to only include the top 50\% producers of original tweets (i.e., tweets that are not replies, quotes, or retweets). 

Before any manual removal of accounts that may have been wrongly included (e.g., if their content is unrelated to the dataset's theme) or that could not be classified into a category (see Supplementary Section 1.1), this selection results in 603 influencers for the COP26 debate, 2168 for the COVID-19 debate, and 1385 for the Ukraine debate.   

\subsubsection*{Manual Annotation}
\label{sec:methods_influencers_annotation}
After identifying a set of influencers for each debate, we manually label each account using ad-hoc questionnaires, in a way similar to~\cite{falkenberg2022growing}. This labeling procedure enables us to associate with each user several features that are not easily retrievable or inferable from their tweets or account information.

We assign each account to one of six predefined categories based on their social, political, or professional role, namely `Activist', `International Organization / NGO', `Media', `Politics', `Private Individual', and `Other'. These categories are shared across all debates. Additionally, we label each account with its main ideological stance in the debate, such as their support for climate action in the COP26 debate, vaccination in the COVID-19 debate, and military aid to Ukraine. We note that the influencer set for each debate is annotated independently of the others, meaning an influencer is labeled with respect to a specific debate only if they appear among the influencers of that particular debate. 

The full list of questions, in addition to further details about the annotation procedure, are reported in Supplementary Section 1.1.

\subsection*{Latent Ideology}
\label{sec:methods_latent_ideology}
The latent ideology technique~\cite{barbera2015tweeting,flamino2023political} allows for estimating a synthetic opinion distribution using retweet interactions. Following a procedure detailed in prior works~\cite{falkenberg2022growing}, we begin by constructing an interaction matrix $M$, where each element $M_{ij}$ is equal to the number of times user $i$ has retweeted user $j$. We exclusively focus on the interactions between users and influencers (as defined in the previous section), with matrix rows corresponding to the former and matrix columns to the latter. Additionally, we ignore all users $i$ such that $\sum_j M_{ij} \leq 1$.
Next, we run Correspondence Analysis \cite{benzécri1973analyse} on matrix $M$. The method requires the following steps. First, we define a new matrix $P$ obtained from the normalization of $M$ with the total number of retweets
\begin{equation*}
P = \frac{M}{\sum_{ij} M_{ij}}
\end{equation*}
and then consider the vector of row sum $\mathbf r = P\mathbf 1$ and column sum $\mathbf c = \mathbf 1^{T} P$. Then, we compute the matrix of standardized residuals of $M$ as
\begin{equation*}
    S = D_r^{-1/2} (P-\mathbf r \mathbf c) D_c^{-1/2}
\end{equation*}
where $D_r =\text{diag}(\mathbf r)$ and $D_c =\text{diag}(\mathbf c)$. Finally, we apply single value decomposition to matrix $S$ as
\begin{equation*}
S = UD_\alpha V^T
\end{equation*}
where $UU^T =VV^T = I$ and $D_\alpha$ is the matrix of singular values of $S$. 
User ideologies are estimated as the standard row coordinates $X =D_r^{-1/2}U$. We focus only on the first dimension, corresponding to the largest singular value, and rescale it to $[-1, +1]$. Like previous works~\cite{flamino2023political, falkenberg2022growing}, influencer ideologies are estimated as the median ideology scores of their retweeters.

\subsection*{Hartigan's dip test}
\label{sec:methods_hartigan}
Hartigan's dip test is a statistical test designed to assess the unimodality or multimodality of a distribution \cite{Hartigan1985}. Specifically, it tests the null hypothesis of an unimodal empirical distribution (i.e., having a single peak) against the alternative hypothesis that it is multimodal (i.e., having more than one peak). The test computes a statistic $D$ that quantifies the extent of multimodality, with larger values corresponding to larger multimodality. The statistical significance $p$ of the test is used to assess the multimodality of the distribution, with low $p$-values indicating that the hypothesis of unimodality can be rejected. Notably, this test can be used to assess whether an empirical ideology distribution can be considered bimodal. In this work, we used this test to assess whether the user opinion distributions inferred with the latent ideology method can be considered bimodal (Fig. \ref{fig:ideology_pdf}), as distributions displaying two peaks at the opposing ends of a spectrum are characteristic of a polarized debate.

\subsection*{Gini Index}
\label{sec:methods_gini}
The Gini Index (or coefficient) is a measure that quantifies the inequality in the frequency values of a distribution \cite{gini1921measurement}. For a vector $\mathbf x$ of $n$ frequencies, it is defined as
\begin{equation*}
    \text{GI} = \frac{\sum_{i=1}^{n} \sum_{j=1}^{n} |x_i - x_j|}{2n^2 \overline{x}}
\end{equation*}
where $\overline{x} = \sum_{i=1}^n x_i / n$ is the mean frequency. The measure is defined in $[0,1]$, with $\text{GI}=0$ corresponding to perfect equality ($x_i=x_j,\; \forall i\neq j$) and $\text{GI}=1$ to perfect inequality ($x_i=1$ and $x_{j\neq i} = 0$). We use this measure to quantify the inequality of the proportion of retweets received by each category, both at the user level and the debate level. 

\section*{Data availability}
Twitter data are made available under the platform's terms of service. Tweet IDs for COP26 are available on OSF (\href{https://osf.io/nu75j/}{https://osf.io/nu75j/}). Tweet IDs for COVID-19 and Ukraine will be made available upon request to the corresponding author.

\section*{Code availability}
The R code used to compute the latent ideology scores is available on OSF (\href{https://osf.io/nu75j/}{https://osf.io/nu75j/}). All other analyses were carried out using base R and freely available packages.

\section*{Author contributions}
A.G. collected the data; E.L., A.G., N.D.M., A.Ba., W.Q. conceived the experiments; E.L. conducted the experiments; E.L., A.G., E.S., N.D.M., M.C., W.Q. analyzed the results; All authors wrote and reviewed the manuscript.

\backmatter

\bmhead{Acknowledgements}
The work is supported by IRIS Infodemic Coalition (UK government, grant no. SCH-00001-3391); SERICS (PE00000014); PRIN 2022 “MUSMA” (CUP G53D23002930006) funded by EU - Next-Generation EU – M4 C2 I1.1; project CRESP from the Italian Ministry of Health under the program CCM 2022; project SEED N. SP122184858BEDB3. A.G. gratefully acknowledges CY4GATE for the financial support.


\end{document}